\documentclass[11pt]{article}

\usepackage{times}

\usepackage{amsmath,amssymb,bbm}
\usepackage{graphicx}
\usepackage[document]{ragged2e}
\usepackage{pagecolor,color}

\newcommand{\noun}[1]{\textsc{#1}}

\usepackage[utf8]{inputenc}
\usepackage[T1]{fontenc}
\usepackage[margin=1.5cm]{geometry}

\usepackage{multicol}
\usepackage{setspace}

\usepackage{natbib}
\usepackage[french]{babel}

\def \draft {1}

\usepackage{xparse}
\usepackage{ifthen}
\DeclareDocumentCommand{\comment}{m o o o o}
{\ifthenelse{\draft=1}{
    \textcolor{red}{\textbf{C : }#1}
    \IfValueT{#2}{\textcolor{blue}{\textbf{A1 : }#2}}
    \IfValueT{#3}{\textcolor{ForestGreen}{\textbf{A2 : }#3}}
    \IfValueT{#4}{\textcolor{red!50!blue}{\textbf{A3 : }#4}}
    \IfValueT{#5}{\textcolor{Aquamarine}{\textbf{A4 : }#5}}
 }{}
}

\newcommand{\todo}[1]{
\ifthenelse{\draft=1}{\textcolor{red!50!blue}{\textbf{TODO : \textit{#1}}}}{}
}



\begin{document}

\title{
Modeling interactions between transportation networks and territories: a co-evolution approach
\bigskip\\
\textit{Journée d'étude Pacte-Citeres 2018\\
}
}
\author{\noun{Juste Raimbault}$^{1,2,3,*}$\medskip\\
$^1$UPS CNRS 3611 ISC-PIF\\
$^2$CASA, UCL\\
$^3$UMR CNRS 8504 Géographie-cités\medskip\\
* \texttt{juste.raimbault@polytechnique.edu}
}
\date{}

\maketitle

\justify


\medskip

\renewcommand{\abstractname}{}
\begin{abstract}
	\begin{center}
	\textbf{Abstract}
	\end{center}
	
	\medskip
	
	Interactions between transportation networks and territories are the subject of open scientific debates, in particular regarding the possible existence of structuring effects of networks, and linked to crucial practical issues of territorial development. We propose an entry on these through co-evolution, and more particularly by the modeling of co-evolution processes between transportation networks and territories. We construct a multi-disciplinary definition of co-evolution which is proper to territorial systems and which can be tested empirically. We then develop the lessons learnt from the development of two types of models, macroscopic interaction models in systems of cities and mesoscopic morphogenesis models through co-evolution. This research opens the perspective of multi-scale models that could be applied to territorial prospective.
	
	\medskip
	
	
\end{abstract}

\section{Introduction}

The potential effect of technical networks on territories, and more particularly of transportation networks, have fed scientific debates which remain relatively open in the current state-of-the-art, such as the issue of identifying of structuring effects of infrastructures~\citep{offner1993effets}. These can be observed on long time periods, but several works recall that caution is key regarding the contingency of each situation and the dangers of a political application of the concept~\citep{espacegeo2014effets}.

A relevant approach to this question, explored by \cite{raimbault2018caracterisation} of which we do a short synthesis here, is to understand territories and transportation networks as co-evolving, i.e. exhibiting strongly coupled dynamics which are difficult to isolate~\citep{bretagnolle:tel-00459720}. The construction of the concept of territory by \cite{raimbault2018caracterisation} at the intersection of Raffestin's approach and Pumain's approach, i.e. by combining the concept of human territory~\citep{raffestin1988reperes} with the one of territory as the space in which systems of cities are embedded~\citep{pumain2018evolutionary}, shows that it implies potential relations between geographical objects, and by induction the emergence of technical networks through the realization of transactional projects~\citep{dupuy1987vers}. Therefore, the interaction between networks and territories can be intrinsically understood as endogenous to the concept of territory itself. We go further by postulating the potential existence of a \emph{co-evolution} between transportation networks and territories, that we will define below from an interdisciplinary point of view in a dedicated section. We study more particularly in our work the \emph{transportation networks}, in particular because these are typically representative of potentialities to shape territories that are attributed to technical networks, more specifically through the effective transportation of flows~\citep{bavoux2005geographie}.

The research work we summarize proposes to explore this perspective of a co-evolution by studying it through the lens of modeling and simulation, considering the model as an instrument of knowledge in itself~\citep{banos2013pour} complementary to the theoretical and empirical aspects~\citep{raimbault2017applied}, and which impact is amplified by the use of methods and tools for model exploration and intensive computation~\citep{pumain2017urban}. The use of generative simulation models for the considered systems allows to construct an indirect knowledge on implied processes and for example to directly test and compare hypothesis in this virtual laboratory~\citep{epstein1996growing}.

Following \cite{raimbault2017invisible} which establishes a map of scientific approaches to the question, domains which studied the modeling of interactions between transportation networks and territories are much varied, from geography to planning of urban economics, and more recently physics, but these seem to be highly isolated~\citep{raimbault2017models}. According to a systematic litterature review and a modelography done by~\cite{raimbault2018caracterisation}, and a typology of interaction processes, two scales appear as relevant to unveil the possible existence of co-evolution processes : the mesoscopic scale, which will typically be a metropolitan spatial scale, and the macroscopic scale, which corresponds to the scale of the system of cities.

Two complementary modeling directions corresponding to these two scales have thus been developed, extending the previous works modeling this co-evolution at the macroscopic~\citep{baptistemodeling,schmitt2014modelisation} and mesoscopic~\citep{raimbault2014hybrid} scale. These two axes correspond to different theoretical frameworks, namely the evolutive urban theory for the macroscopic \citep{pumain2018evolutionary} and urban morphogenesis for the mesoscopic, which we understand as the emergent relation between form and function~\citep{doursat2012morphogenetic}.

This contribution aims at giving a synthesis of these research works, and is organized the following way : we firstly define the concept of co-evolution from a multi-disciplinary viewpoint. We then detail the results obtained for the modeling at the macroscopic scale, and then the ones for the mesoscopic scale. We finally discuss the perspectives opened and the future developments in the context of modeling co-evolution between transportation networks and territories.

\section{Defining co-evolution}

Transfer of concepts between disciplines is always difficult, and as co-evolution has been used by several disciplines, we propose here a definition inspired by the various ones in which it was developed. A multi-disciplinary review and a more general definition not specific to territorial systems is developed in \cite{raimbault2018co}. The detailed description of the different approaches is given in this paper, of which we give here only a broad summary.

Originating in biology, the concept of evolution requires typical features for a system to exhibit it~\citep{durham1991coevolution}, namely (i) transmission processes between agents; (ii) transformation processes; and (iii) isolation of sub-system such that differentiations emerge from the previous processes. Co-evolution then corresponds to entangled evolutionary changes in two species~\citep{janzen1980coevolution}. It was generalized to diffuse co-evolution by taking into account the broader context of numerous species in the ecological interaction network and of the environment~\citep{strauss2005toward}.

The concept was transferred to disciplines closer to social and human sciences, including the emerging field of cultural evolution \cite{Mesoudi25072017} for which building bricks of culture are transmitted and mutated, possibly with an interplay with biological evolution itself~\citep{bull2000meme}, but also sociology to interpret for example the interactions between social organizations as entities themselves~\citep{volberda2003co}. In the frame of evolutionary economics~\citep{nelson2009evolutionary}, the concept was also largely applied in economic geography~\citep{schamp201020}, to investigate for example the link between economic clusters and knowledge spillovers~\citep{doi:10.1080/00343400802662658}, the link between territories and technological innovation~\citep{colletis2010co}, or environmental economics issues~\citep{kallis2007coevolution}. The concept was particularly developed in geography in the frame of the evolutive urban theory \citep{pumain1997pour}. Its operational definition in that context taken for example by \cite{paulus2004coevolution} or \cite{schmitt2014modelisation} relies on systems of cities constituted by strongly coupled subsystems and entangled interactions.

We can observe that these various definitions of co-evolution mostly correspond to the theoretical framework introduced by \cite{holland2012signals}, in which hierarchically imbricated subsystems correspond to ecological niches and are therefore containing co-evolutive dynamics.


The definition of co-evolution we construct for the particular case of transportation networks and territories is the following, similar to the definition of \cite{raimbault2018co}:

\begin{enumerate}
    \item the evolutive processes are carried by the transformation of territorial components at different scales;
	\item a co-evolution may occur in the strong coupling of such evolutive processes, with different levels of strength, namely as circular causal relation (i) between individual entities; (ii) within populations of individual, i.e. with a certain statistical sense within a given territory; (iii) between most elements of the system with a difficulty to disentangle these relations.
	\item these different level and the spatial isolation typically enhancing evolutionary drift imply the existence of \emph{territorial niches}, i.e. niches of co-evolution, imbricated at different scales.
\end{enumerate}

Our definition is strongly anchored within a dynamical vision of processes, within the initial spirit of the introduction of the concept in biology. The opening on the different levels to which the co-evolution can occur yields a generality but also a precision and the construction of empirical characterization methods, such as the one introduced for the second level (population co-evolution) by \cite{raimbault2017identification}. Furthermore, the implicit integration of the concept of niche implies a suitability with territoriality and its declination at different scales within territorial subsystems both independent and interdependent. Our approach is more general than the notion of congruence proposed by \cite{offner1993effets}, which remains fuzzy in the interdependency relations between the entities concerned, and could be similar to the third level of systemic interdependencies.

\section{Macroscopic scale}

The first modeling axis relates to the macroscopic scale and is based on the principles of the evolutionary urban theory~\citep{pumain1997pour}. The family of Simpop models is mainly situated in corresponding ontologies and scales, i.e. elementary entities constituted by cities themselves, at the spatial scale of the system of cities (regional to continental) and on relatively long time scales (longer than half a century)~\citep{pumain2012multi}.

\subsection{Network effects}

A first model that can be interpreted as a control, in which the network is static but has a retroaction on cities, indirectly suggests network effects. This preliminary work is detailed by~\cite{raimbault2018indirect} which details the model and applies it to the French system of cities on a long time scale (1830-1999). The model is based on expected populations and captures complexity through non-linear interactions between cities, which are carried by the network. Three processes are added to determine the growth rate of cities: (i) an endogenous growth fixed by a parameter, corresponding to the Gibrat model ; (ii) direct interaction processes described as a gravity potential which influences the growth rate ; (iii) a retroaction of flows circulating in the network on the cities traversed. The model is initialized with real populations at the start of a period, and then evaluated by comparison to simulated populations on the full period, on two objectives which allow to take into account the adjustment of the total population or of their logarithm. The production of Pareto fronts shows that it is not possible to uniformly adjust the model for the full spectrum of city sizes. We show that the addition of the network component provides an effective increase in the adjustment, what suggests network effects.

\subsection{Co-evolution model}

This model is then extended to a co-evolutive model by~\cite{raimbault2018modeling}, in which cities and links of the transportation network are both dynamic and within a reciprocal dependency. This model is close to~\cite{schmitt2014modelisation} for the rules of population evolution, and to~\cite{baptistemodeling} for the evolution of the network. More precisely, it operates iteratively with the following steps : (i) population of cities evolve according the the specification of the static model described above ; (ii) the network evolves, following an abstract implementation such that distances between cities are updated with a self-reinforcement function depending on flows between each city, with a threshold parameter. This version of the model is strictly macroscopic and does not include the spatial form of the network since it updates the distance matrix only.

The systematic exploration of this model using the OpenMOLE software~\citep{reuillon2013openmole} and the application of an empirical method to characterize co-evolution~\citep{raimbault2017identification} allow us to show that it captures a large variety of coupled dynamics, including indeed co-evolutive dynamics: among the broad range of interaction regimes between network variables and population variables (in the sense of \cite{raimbault2017identification}, by classifying lagged correlation patterns), more than half of the regimes are effectively co-evolutive, i.e. exhibit circular causalities. This aspect could appear as trivial for a model conceived to integrate a co-evolution, but one has to realize that processes integrated in models are at the microscopic scale whereas the co-evolution is quantified at the macroscopic scale: it is indeed a property of the model to make co-evolution emerge. For example in the case of the SimpopNet model~\citep{schmitt2014modelisation}, \cite{raimbault2018unveiling} shows that co-evolution regimes are much more rare and that this other model produces more often configurations of the type structuring effects or without any relation. Furthermore, the exploration unveils the existence of an optimal value for an interaction range parameter, at which the system exhibits a maximal complexity of city trajectories. This range correspond to the appearance of territorial niches, within which populations are co-evolving, corresponding to the third point of our definition.

The calibration on the French system of cities on the same time period than the static model, with population data and dynamical railway network data taken into account with dynamical distance matrices constructed from the database of \cite{thevenin2013mapping}, unveils Pareto fronts between the objective on the distance between cities and population objective, suggesting an impossibility to simultaneously calibrate this type of models for the network component and for the territorial component. Moreover, the adjustment on populations is improved by this model for a certain number of periods, compared to the model with a static network, suggesting the relevance of taking into account co-evolutive dynamics. The evolution of the calibrated threshold parameter suggests that the model captures a ``High Speed Rail (TGV) effect'', i.e. an increase in the effective accessibility for the metropolitan areas concerned but a loss of speed for the territories left behind.

\section{Mesoscopic scale}

The second axis, at the mesoscopic scale, considers the approach through urban morphogenesis, understood as the simultaneous emergence of the form and the function of a system \citep{doursat2012morphogenetic}. It allows to consider a more precise description of territories, at the scale of fine population grids (500m resolution) and of a vectorial representation of the network at the same scale.

\subsection{Morphogenesis by aggregation-diffusion}

The territorial systems produced are quantified with morphological indicators for the population \citep{le2015forme}. A first preliminary model, integrating only the population, shows that aggregation and diffusion processes are sufficient to explain a large majority of urban forms existing in Europe \citep{raimbault2018calibration}. This result suggests that taking into account the form only can be achieved in an autonomous way, but that functional processes will then not be taken into account in the core of dynamics. In order to grasp morphogenesis processes, i.e. the strong link between form and function during the emergence of these, we follow the idea of using the transportation network as a proxy of functional properties of territories, in particular through centrality properties. This leads us to consider a morphogenesis model by co-evolution at the mesoscopic scale.

\subsection{Morphogenesis by co-evolution}

The urban form indicators computed on moving windows of size 50km for the whole Europe are completed with structural network indicators for the road network, computed from OpenStreetMap data after application of a specific simplification algorithm which conserves topological properties \citep{raimbault2018urban}. These indicators and their static spatial correlations are thus computed on windows of a similar size covering the whole Europe. We show through the study of the spatial study of these correlations the non-stationarity of interaction processes at the second order, confirming the relevance of the notion of niche as a consistent territorial sub-system.

We then introduce in \cite{raimbault2018urban} a morphogenesis model capturing the co-evolution of the spatial distribution of population and of the road network. This model combines the logic of \cite{raimbault2018calibration} for the complementarity of aggregation and diffusion processes to the one of \cite{raimbault2014hybrid} for the hybrid grid and vectorial network structure and also the influence of local explicative variables on the territorial evolution. This model works the following way: (i) morphological and functional local properties, integrated as local normalized explicative variables (including population, distance to the network, betweenness centrality, closeness centrality, accessibility), determine the value of a utility function from which new population is added through preferential attachment, and a diffusion of population is achieved; (ii) the road network evolves following rules depending on different heuristics (multi-modeling approach), which include, after the addition of nodes preferentially to the new population and their direct connection to the existing network, an addition of links with an heuristic among random links, random potential breakdown \citep{schmitt2014modelisation}, deterministic potential breakdown \citep{raimbault2016generation}, biological self-organized network \citep{tero2010rules}, cost-benefit compromises \citep{louf2013emergence}.

The computation of topological indicators for the road network allows to calibrate the model, and \cite{raimbault2018multi} shows that the different network growth processes which have been included following the multi-modeling procedure are complementary to reach a maximum a real network configurations. Furthermore, the model is simultaneously calibrated on morphological indicators, topological indicators, and their correlation matrices, and the relatively low distances to data for a non negligible number of points suggest that the model is able to reproduce outcomes of processes at the first order (indicators) but also at the second order (interactions between indicators). Regarding the causality regimes produced by the model, we obtain configurations corresponding to a co-evolution, but a much lower diversity than for the more simple model of \cite{raimbault2014hybrid} which was used as a toy model to show the relevance of the method of causality regimes in \cite{raimbault2017identification}, suggesting a tension between static performance (reproduction of outcomes of processes) and dynamical performance (reproduction of processes themselves) for this kind of models.

\subsection{Transportation governance}
 
Finally, a last metropolitan model (Lutecia model) is described in \cite{raimbault2018caracterisation}, extending the one proposed by \citep{lenechet:halshs-01272236}. It explores the role of governance processes in the growth of the transportation network, within a co-evolution model. Here, the metropolitan scale indeed corresponds to the mesoscopic scale. The collaboration between local actors for the construction of transportation infrastructures is included using game theory paradigms. This allows to simulate the emergence of the transportation network and its interaction with the urban form quantified by spatial accessibility patterns. This model allows for example to show that co-evolutive dynamics can in some case lead to the inversion of the behavior of accessibility gains in comparison to a situation without evolution of land-use, i.e. to qualitatively change the regime of the metropolitan system. The calibration of this model on the stylized case of the mega-urban region of Pearl River Delta in China allows to extrapolate on governance processes, and suggests that the shape of the current network is more likely to be the consequence either of fully regional decisions or of fully local decisions, but never an intermediate configuration, contradicting the view of the Chinese political context as a multi-level decision system~\citep{liao2017ouverture}.

\section{Perspectives}

This research therefore developed complementary approaches at different scales of interactions between transportation networks and territories by modeling their co-evolution. We finally detail some immediate development perspectives opened by this work.

\subsection{Developments at the macroscopic scale}

Our macroscopic models have not been tested yet on other urban systems and other time spans, and future developments will have to study which conclusions obtained here are specific to the French urban system on these periods, and which are more general are could be more generic in systems of cities. The application of the model to other systems of cities also recalls the difficulty to define urban systems. In our case, a strong bias must be induced by the fact of considering France only, since the insertion of its urban system within an European system is a reality we had to neglect. The span and scale of such models is always a difficult subject. We rely here on the administrative consistence and on the consistence of the database \citep{pumain1986fichier}, but the sensitivity to the definition of the system and to its spatial extent must still be tested.

Furthermore, the calibration used only the railway network for the distances between cities. Considering a single transportation mode is naturally a reduction, and an immediate direction for developments is testing the model with real distance matrices for other types of networks, such as the freeway network which has undergone a significant growth in France in the second half of the 20th century. This application would necessitate the construction of a dynamical database for the freeway network spanning 1950-2015, since classical bases (IGN or OpenStreetMap) do not integrate the opening date of segments. A natural extension of the model would then consist in the implementation of a multilayer network, what is a typical approach to represent multi-modal transportation networks~\citep{gallotti2014anatomy}. Each layer of the transportation network should have a co-evolutive dynamic with populations, possibly with the existence of inter-layer dynamics.

\subsection{Developments at the mesoscopic scale}

The issue of the generic character of the co-evolution morphogenesis model is also opened, i.e. if it would work similarly to reproduce urban forms on very different systems such as the United States or China. A first interesting development would be to test it on these systems and at slightly different scales (1km cell size for example).

Furthermore, the Lutecia model is also a fundamental contribution towards the inclusion of more complex processes implied in co-evolution, such as the governance of the transportation system. This model paves the way to a new generation of models, that could potentially become operational in the case of regional systems with a very high evolution speed such as in the Chinese case.

\subsection{Towards multi-scale models}

Our work finally opens perspectives for integrated approaches, towards multi-scale models of these interactions, which appear to be more and more necessary for the construction of operational models that can be applied to the design of sustainable planning policies \citep{rozenblat2018conclusion}.

A first contribution towards multi-scalar models would be to take into account in a finer way the physical network in macroscopic models, what is for example the object of~\cite{mimeur:tel-01451164}, which produces interesting results regarding the influence of the centralization of network investment decision-making on final forms, but keeps static populations and does not produce a co-evolution model. Similarly, the choice of indicators to quantify the distance of the simulated network to a real network is a difficult question in that context: indicators such as the number of intersections taken by~\cite{mimeur:tel-01451164} is associated to procedural modeling and does not reflect structural indicators. It is probably for the same reason that~\cite{schmitt2014modelisation} is only interested in population trajectories and not in network indicators: the superposition and adjustment of population and network dynamics at different scales remains an open problem.

A second entry consists in the integration of the mesoscopic morphogenesis model into population of models in interaction. This approach would allow to take into account the non-stationarity of territorial systems that we moreover showed empirically. Here at the mesoscopic scale, total population and growth rate are fixed by exogenous conditions due to processes at the macroscopic scale. It is in particular the aim of spatial growth models such as the macroscopic model we already introduced to determine such parameters through the relations between cities as agents. It would then be possible to condition the morphological development of each area to the values of parameters determined at the upper level. In that context, one must remain careful on the role of emergence: should the emergent urban form influence the macroscopic behavior in its turn ? Such complex multi-scalar models are promising but must be considered with caution for the level of complexity required and the way to couple scales.

\section{Conclusion}

We did here a synthesis of main results of \cite{raimbault2018caracterisation}, confirming the relevance of the co-evolution approach to focus on interactions between transportation networks and territories, in particular to model these interactions. We developed a precise definition of co-evolution, moreover associated to an empirical characterization method. The models at different scales, and the perspective of multi-scale models, can become precious tools for territorial prospective in the context of sustainable transitions on long time, allowing to quantify the possible territorial systems and the one desirables regarding sustainability, taking into account scales and processes still relatively poorly tackled in the literature.





\end{document}